# Ambipolar charge injection and transport in a single pentacene monolayer island

T. HEIM, K. LMIMOUNI & D. VUILLAUME[1]

Institut d'Electronique, de Micro-electronique et de Nanotechnologie – CNRS

BP 69, avenue Poincaré, F-59652cedex, Villeneuve d'Ascq, France.

vuillaume@isen.iemn.univ-lille1.fr

ABSTRACT. Electrons and holes are locally injected in a single pentacene monolayer island. The two-dimensional distribution and concentration of the injected carriers are measured by electrical force microscopy. In crystalline monolayer islands, both carriers are delocalized over the whole island. On disordered monolayer, carriers stay localized at their injection point. These results provide insight into the electronic properties, at the nanometer scale, of organic monolayers governing performances of organic transistors and molecular devices.

---

[1] Corresponding author : vuillaume@isen.iemn.univ-lille1.fr

Organic field effect transistors (OFET), using a thin film of molecules or polymers as the semiconductor layer, have gained an increasingly interest for low cost, lightweight, flexible and large area electronics.[1] Charge carrier mobility µ is now in the order of ~ 0.1 $cm^2V^{-1}s^{-1}$ for polymer OFET and of about a few $cm^2V^{-1}s^{-1}$ for OFET made of sublimated films of small oligomers (e.g. pentacene) and single crystals. It has been long recognized[2] that the transport properties in the OFET strongly depends on the morphology and molecular structure of the first few molecular layers deposited on the gate dielectric (mainly $SiO_2$). Electrostatic principles clearly predict that the charge transporting layer is confined close to the gate dielectric interface, within the first few organic monolayers.[3] In this respect, several recent works[4-6] have dealt with detailed analysis of the organic film growth kinetics, on the related growth mechanisms and the relationship with the chemical and physical properties of the gate dielectric surfaces. However, only scarce results have been so far published on the dependence of the transport properties with the organic semiconductor film thickness.[2,7] Recently, Dinelli and coworkers[8] reported that 2 organic monolayers (sexithiophene molecules) are requested to obtain a hole field-effect mobility on a par with those of OFET based on thicker organic films. Below 2 monolayers, the hole mobility strongly decreases by about 2 order of magnitude. This result is similar to our own result that an OFET based on 3 monolayers of pentacene have the same mobility than the OFET with a 50 nm (~ 30 monolayers) pentacene film.[9] However, these results have been obtained at a scale of several tens of micrometers corresponding to the lateral distance between the source and drain electrodes. As a consequence, grain boundaries, that strongly limit the mobility and the charge transport in organic films,[10] are still playing a major role in these experiments. To establish the ultimate performances of these organic materials and transistors, it is mandatory to measure the transport properties of a single organic domain with a monolayer thickness.

In this Letter, we report how electrons and holes, that are locally injected in a single pentacene monolayer island, stay localized or are able to delocalize over the island as a function of the molecular conformation (order vs. disorder) of this island. Charge carriers were locally injected by the apex of an atomic force microscope tip, and the resulting two-dimensional distribution and concentration of injected charges were measured by electrical force microscopy (EFM) experiments. We show that in crystalline monolayer islands, both electrons and holes can be equally injected, at a similar charge concentration for symmetric injection bias conditions, and that both charge carriers are delocalized over the whole island. On the contrary, charges injected into a more disordered monolayer stay localized at their injection point. These results provide insight into the electronic properties, at the nanometer scale, of organic monolayers governing performances of organic transistors and molecular devices.

We prepared the isolated islands of pentacene by high vacuum ($10^{-7}$–$10^{-6}$ torr) sublimation in the sub-monolayer coverage regime. We used two different types of substrate. The sample **I** is made on a n-type silicon wafer, covered by ~ 4 nm thick thermal oxide. The $SiO_2$ surface roughness (rms) is ~ 0.15 nm (AFM measurements). The $SiO_2$ surfaces were cleaned in acetone, isopropyl alcohol and de-ionized water before the pentacene deposition. The pentacene was sublimated on the substrate kept at 293 K (20°C) and at a very low deposition rate of $2.5 \times 10^{-3}$ Å·s$^{-1}$. We used the pentacene as received without further purification. We stopped the evaporation at an average nominal thickness (as given by the quartz crystal balance of the evaporator) ~0.3 nm (about 20 min. of deposition) corresponding to a sub-monolayer coverage. For the sample **II**, we used a n-type silicon wafer which is now covered with a rough native oxide, the roughness is ~ 0.4 nm. The pentacene deposition rate was $3.3 \times 10^{-2}$ Å·s$^{-1}$. We deliberately used rougher oxide (roughness of about one third of the pentacene length) and higher deposition rate to obtain a more disordered pentacene monolayer than in sample **I**.[11] The pentacene evaporation was stopped after about 10 s (average thickness not detectable with the quartz crystal balance). Tapping mode AFM (TM-AFM) images (Fig. 1) show that we have obtained isolated islands of pentacene. On substrate **I**, the islands have a dendritic geometry. The average island size is ~ 1-1.3 µm$^2$ with a density of about 20 islands per 100 µm$^2$ and thus a surface coverage θ ~ 0.2. The form factor of these islands, $4\pi \langle A \rangle / \langle P \rangle^2$ where <A> is the mean island area and <P> the mean perimeter, is 0.11 and their fractal dimension[12] is d ~ 1.54 in agreement with other reports.[4-6] These values correspond to a diffusion limited aggregation growth mechanism. On substrate **II**, we have obtained smaller islands (typical average size of ~ 0.1 µm$^2$) at about the same density (~ 20 islands per 100 µm$^2$) corresponding to a lower coverage θ ~ 0.02. In both samples, the profile sections (see Figs. 2-a and 4-a for typical exemples) show small island-to-island thickness variations. All measured thickness are in the range 1.5 to 1.9 nm, which corresponds to one monolayer of pentacene molecules standing more or less upright along the normal of the substrate surface (pentacene length along the main axis is 1.64 nm). Roughness measurements on the island surface show that the pentacene island surface has about the same roughness[13] as the underlying substrate (~ 0.2 nm for pentacene island on sample **I**, and ~0.6 nm for island on substrate **II**, compared to ~0.15 and ~0.4 nm for the SiO$_2$ substrate, respectively – see above). Local charge injections and EFM experiments were performed using a Nanoscope III microscope (Digital Instruments) under dry nitrogen atmosphere. We used PtIr-coated cantilevers with a free oscillating frequency $f_0$ ~ 60 kHz and a spring constant k ~ 1-3 N/m. To locally inject charges into the pentacene monolayer islands, the EFM tip was biased at $V_{INJ}$ with respect to the silicon wafer, its oscillation frequency was set to zero and the tip was gently contacted to the pentacene island with a typical 2nN contact force for a few seconds to minutes (Fig. 1). The 2D-distributions of charges injected in the pentacene monolayer island were then characterized by EFM, in which electric force gradients acting on the tip biased at $V_{EFM}$ shift the EFM cantilever phase (or oscillation frequency). For this measurement, the tip-substrate distance was typically in the range 50-100 nm. EFM images reveal two distinct interactions. First, the capacitive interaction associated with the local increase of the tip-substrate capacitance when the EFM tip is moved over the pentacene monolayer island leads to negative phase shifts varying as $-V_{EFM}^2$ (see dark features in Fig. 2-b). The second interaction is the interaction between the charge Q stored in the pentacene monolayer island and the capacitive charge at the tip apex. This additional phase shift is either positive or negative, and varies as $QxV_{EFM}$. When $QxV_{EFM} > 0$ (repulsive interaction), that corresponds to a positive phase shift, leading to bright features in the EFM phase images. On the contrary, $QxV_{EFM} < 0$ (attractive interaction) corresponds to a negative phase shift and thus a dark features in the phase images. Capacitive and charge interactions can be distinguished by measuring the phase shift as a function of $V_{EFM}$. To quantify the amount of charge in the pentacene islands, we used a recently proposed analytical model.[14] The capacitive force gradient leads to a phase shift expressed by $\Delta\Phi_C = -(f_0/2k)3\varepsilon_0 Sh(V_{EFM}-V_S)^2/z^4$ where $\varepsilon_0$ is the vacuum permittivity, S the area of the capacitance between tip and substrate, h the island height and z the tip-substrate distance and $V_S$ the surface potential. The phase shift due to charge force gradient is given by $\Delta\Phi_Q = (f_0/2k)Qh(V_{EFM}-V_S)/\varepsilon_R z^3$ with Q the stored charge in the island and $\varepsilon_R$ the pentacene dielectric constant. According to the model and protocol developed elsewhere,[14] we calculated the charge Q from the ratio $R=\Delta\Phi_Q/\Delta\Phi_C$, this ratio does not depend on peculiar properties of the cantilever and EFM setup (spring constant, quality factor, lateral resolution,…) used in the experiments. It has been demonstrated,[14] by a careful comparison with numerical simulations, that this ratio, meanwhile analytically derived from a plane capacitor geometry between the substrate and the tip, remains valid by introducing two correction factors to take into account the island and tip shapes. If we express the stored charge as a surface density, $\sigma$, in the pentacene monolayer island, $Q=e\sigma S$ (e is the electron charge), we get:[14]

$$R = -\frac{g}{\alpha}\frac{(\sigma/\varepsilon_R)ez}{\varepsilon_0(V_{EFM}-V_S)} \quad (1)$$

The tip shape factor, $\alpha$, is ~1.5 for a standard conical tip. The island shape factor, g, is about 1 if we approximate the pentacene island as a disk. This analytical model has been derived for a nano-object deposited onto a metallic substrate. Here, we have to take into account the ultra-thin oxide (2 to 4 nm) between the pentacene monolayer island and the silicon substrate.[15] Actually, the form factor is g ~ 5 with a 4 nm thick oxide (sample **I**) and g ~ 3 with a 2 nm thick oxide (sample **II**).[16]

Figure 2 shows the topographic AFM image of the pentacene monolayer islands (sample **I**) and the EFM images before and after the charge injection. Before the charge injection, all the islands appear as slightly dark in the EFM image (Fig. 2-b). These dark features are due to the capacitance coupling effect and/or due to a small residual charge in the islands. Charge and capacitive coupling effects were carefully separated by measuring the relationship between the EFM phase shift and $V_{EFM}$. Figure 3 shows a typical example for the pentacene monolayer island of the sample **I** before any charge injection.

The $\Delta\Phi$-$V_{EFM}$ curve varies as $V_{EFM}^2$ but the curve is slightly shifted by a linear component (fit in Fig. 3). This reveals that the as-grown pentacene islands are slightly positively charged. From the ratio between the charge contribution (the term varying as $V_{EFM}$) and the capacitive contribution (the term varying as $V_{EFM}^2$), we estimated from a second order polynomial fit of the $\Delta\Phi$-$V_{EFM}$ curve in Fig. 3 an effective surface charge density ($\sigma/\varepsilon_R$) of the pentacene island of ~200 charges/µm². If we take a relative index $\varepsilon_R \sim 3$ for the pentacene and if we consider that one molecule of pentacene in the monolayer occupies an area of ~ 22.5 Å², this value corresponds to ~ $1.3 \times 10^{-4}$ charge per pentacene molecule (the area occupied on the surface by a molecule of pentacene is deduced from the unit cell measured by grazing-angle incidence X-ray diffraction[17] on a pentacene monolayer). Converted in a bulk unit, this value represents a residual doping of the pentacene semiconductor of about $3.7 \times 10^{14}$ cm⁻³. We injected the charges in the center location of one of the islands (indicated by the arrow in Fig. 2). The injection conditions where $V_{INJ}=+2V$ for $t_{INJ}=4$ min. Figure 2-c shows the EFM image (still at $V_{EFM}= + 4V$) after this injection. From the large bright feature in the EFM image, it is clear that the injected charges are delocalized along the entire pentacene island. No change is seen in this EFM image for the other neighboring islands on which we did not inject any charge. The bright feature indicates that $Q \times V_{EFM} > 0$ and thus that a net positive charge is now stored in the island. This is consistent with the fact that holes are easily injected from the Pt/Ir tip (work function of -5.4 eV) in the pentacene HOMO (-5 eV) when the tip is positively bias with respect to the n-Si substrate. On the contrary, electrons have to overcome the energy barrier of the 4 nm thick oxide to be injected from the Si substrate. This process is much less probable. From the amplitude of the EFM signals (measured from the cross-sections shown in Figs. 2-b and 2-c) after and before injections, we calculated the ratio R ~ -10 (to estimate $\Delta\Phi_C$ we assumed that the EFM signal before injection is mainly due to capacitive coupling, neglecting the residual small charge discussed before). In that case, $\Delta\Phi_Q \sim 4°$ (Fig. 2-c) and $\Delta\Phi_C \sim -0.4°$ (Fig. 2-b). We deduced, using Eq. (1), an effective charge $\sigma/\varepsilon_R$ of $6.6 \times 10^3$ charges/µm² or ~ $4.5 \times 10^{-3}$ hole per pentacene molecules (with the same hypothesis as before on the value of $\varepsilon_R$ and of the area per pentacene molecule) and an equivalent bulk concentration of ~ $1.2 \times 10^{16}$ cm⁻³. Then, we again put the tip on the island center and submitted the same pentacene monolayer island at $V_{INJ} = -2V$ for 4 min. Fig. 2-d shows the EFM image recorded at $V_{EFM} = +4V$ after this experiments. The strong dark feature indicates that now $Q \times V_{EFM} < 0$ and thus that a net negative charge is now stored in the pentacene monolayer island. This demonstrates that previously injected holes have been removed (or compensated), and that excess negative charges are now totally delocalized over the entire domain. Using the same protocol as before, the amplitudes of the capacitive and charge EFM signals give R = 12, and we deduced an effective charge $\sigma/\varepsilon_R$ of $-9.8 \times 10^3$ charges/µm$^2$ (or ~ $6.6 \times 10^{-3}$ electron per pentacene molecules with the same assumption as before and an equivalent bulk concentration of ~ $1.8 \times 10^{16}$ cm$^{-3}$). Given the energetic of pentacene, at equilibrium (null bias) the LUMO is at about 2.2 eV above the Fermi energy of the Pt/Ir tip. From the Gauss law, we calculated that at the injection bias of 2 V only ~ 1.05 V are applied to the pentacene monolayer, the rest being lost in the oxide (this value takes into account the applied field and the internal field due to the work function difference between Pt/Ir and n-Si). This potential is not sufficient to align the LUMO of pentacene and the Fermi energy of the tip so as allowing an easy electron injection from the tip. However, these numbers are based on energy levels known for bulk pentacene and the present case of one pentacene monolayer sandwiched between SiO$_2$ and metal need to be examined carefully. Our results show that about the same amount of holes and electrons can be injected in the pentacene monolayer island at symmetric bias injection conditions and that both types of carriers easily delocalize over the whole pentacene monolayer. This ambipolar behavior suggests that at the tip/pentacene interface, the Pt/Ir Fermi energy is probably located at about half of the pentacene monolayer HOMO-LUMO gap. It has been recently demonstrated[18] that there is no significant change of energy level alignment at the pentacene-SiO$_2$ interface, but an interfacial dipole certainly exist at the Pt/Ir-pentacene interface as it is well-known for other metal/organic systems.[19] To take into account the fact that electrons are injected as easily as holes, we assume that this interface dipole induces an energy shift of ~ 1 eV so as to lower the pentacene molecular orbitals with respect to the Pt/Ir Fermi energy.[20] A small halo of positive charges remains at the border of the pentacene monolayer island (Fig. 2-d). Possible reasons may be: i) a smaller diffusion of electrons than holes; ii) a different 2D distribution of field and potential between the tip and the substrate when the pentacene island is uncharged (as it is the case for the first injection at $V_{INJ} = 2V$) and when the pentacene island is positively charged (as it in the case for the second injection at $V_{INJ} = -2V$), iii) trapping of the injected holes at border defects, these trapped holes may be not easily detrapped or recombined during the electron injection. The first hypothesis may be discarded because a diffusion length $l_D$ limited to ~ 1µm (i.e. the mean size of the pentacene island) would correspond to a mobility $\mu = el_D^2/kT\tau$ of the order of $10^{-9}$ cm$^2$V$^{-1}$s$^{-1}$ (taken a typical time constant of 280 s, i.e. the injection time plus the EFM image measurement time of ~ 40 s). This value is well below the expected mobility (compared to usual pentacene films)[1] and the mobility is probably not the limiting factor. However, the mobility in a monolayer might be strongly different from that in a more bulky film and the exact measurement of hole and electron diffusions/mobilities in our pentacene monolayer islands require a more detailed dynamic study of the charge delocalization at variable shorter injection times and also on larger pentacene monolayer islands. A 2D simulation of field and potential distribution between the tip and the pentacene island is required to confirm the second hypothesis. Both works are beyond the scope of this letter, they are in progress and will be reported latter. The general behavior reported here for sample **I** is reproducible and was observed for 3 different pentacene islands. Inverting the injection sequence (electrons first followed by holes) gave similar EFM observations (see supporting information).

We repeated the same series of injection/detection experiments on the pentacene monolayer islands of the sample **II** (Fig. 4). Before any injection (Fig. 4-b), $\Delta\Phi$ is not always negative as expected when the capacitive force gradient dominates (as shown in the case of sample **I**, Fig. 3). Here, $\Delta\Phi > 0$ for $V_{EFM} = -5V$ and $\Delta\Phi < 0$ for $V_{EFM} = +5V$. This means that a larger residual negative charge stored in the pentacene monolayer island dominates the force gradient. We estimated this residual charge $\sigma/\varepsilon_R$ ~ -1.7x10$^4$ charges/µm$^2$, i.e. ~ 1.1x10$^{-2}$ negative charge/molecule (or a bulk residual doping of ~ 3x10$^{16}$ cm$^{-3}$).[21] We injected holes from the tip at $V_{INJ} = +5V$ for 30 s. We used a smaller injection time (compared to sample **I**) to keep the total time of the experiment (injection and EFM image measurement, ~ 40-50 s for the latter) smaller than the typical charge retention time in the pentacene island. This is to ensure a correct determination of the stored charge. The retention time is fixed by charge leakage through the oxide layer between the pentacene island and the silicon substrate. Thinner the oxide is, smaller the retention time.[22] A retention time of about few minutes has been inferred for nanoparticle capped with a ~ 2 nm thick oxide.[23] However, to inject a comparable amount of charges (~ $10^4$ charges/µm$^2$) we did the injection at higher voltage than in sample **I**. After the injection we did not see any delocalization of the injected holes (Fig. 4-c, measured at $V_{EFM}$ = +5V). We only observed a bright spot at the injection point. The $\Delta\Phi$ profile in Fig. 4-c shows that i) almost all the residual negative charges have been compensated at this point by the injected hole (the ration of the $\Delta\Phi$ peak height over the $\Delta\Phi$ value at the bottom in the island gives ~ 70%), ii) the full width at half maximum (FWHM) of this $\Delta\Phi$ peak is ~ 68 nm. This FWHM of the localized charges is fully consistent with the tip radius (15-20 nm) used for both injection and detection and with the typical EFM lateral resolution of ~ 50 nm in our experimental conditions.[24]

The transport properties in organic monolayers strongly depend on the π-π orbital overlap between neighboring molecules, which is very sensitive to the molecular packing within the monolayer.[25] It was demonstrated recently, by grazing-angle incidence X-ray diffraction, that a pentacene monolayer island grown at a low rate on a very flat state-of-the-art thermal oxide (as it is the case on substrate **I**) is crystalline.[17] The charge delocalization observed in our injection and EFM experiments on sample **I** is consistent with this good molecular packing in the island, and it demonstrates that a good π-π overlap exists in such a pentacene monolayer. On the contrary, the pentacene monolayer deposited on substrate **II** is certainly much more disordered since a surface with a roughness of one third of the molecule length (roughness is 0.4 nm for substrate **II**) can destroy or at least strongly disturb the molecule packing in the monolayer. Our results on such pentacene monolayers that the injected charges are not delocalized over the whole monolayer are consistent with this hypothesis.

In conclusion, these results demonstrate that the local injection of charges in a single pentacene monolayer island, followed by the EFM detection of 2D distribution of charge to observe their delocalized or localized behavior, is a useful technique to study the electronic properties of such organic films at an early stage of growth. In addition, the method gives access to a quantitative determination of the amount and the map of the injected charges, as well as to the residual doping of the as-grown films. These results provide insight into the electronic properties, at the nanometer scale, of the organic monolayers playing a major role in organic transistors. This injection/EFM detection technique can also be used to characterize other monolayers and molecules of interest for molecular-scale electronics.[26].

ACKNOWLEDGEMENT. We thank T. Melin for useful discussions and D. Deresmes for technical help with the AFM-EFM equipment. The work was partly supported by IRCICA (Institut de Recherche sur les Composants logiciels et matériels pour l'Informatique et la Communication Avancée).

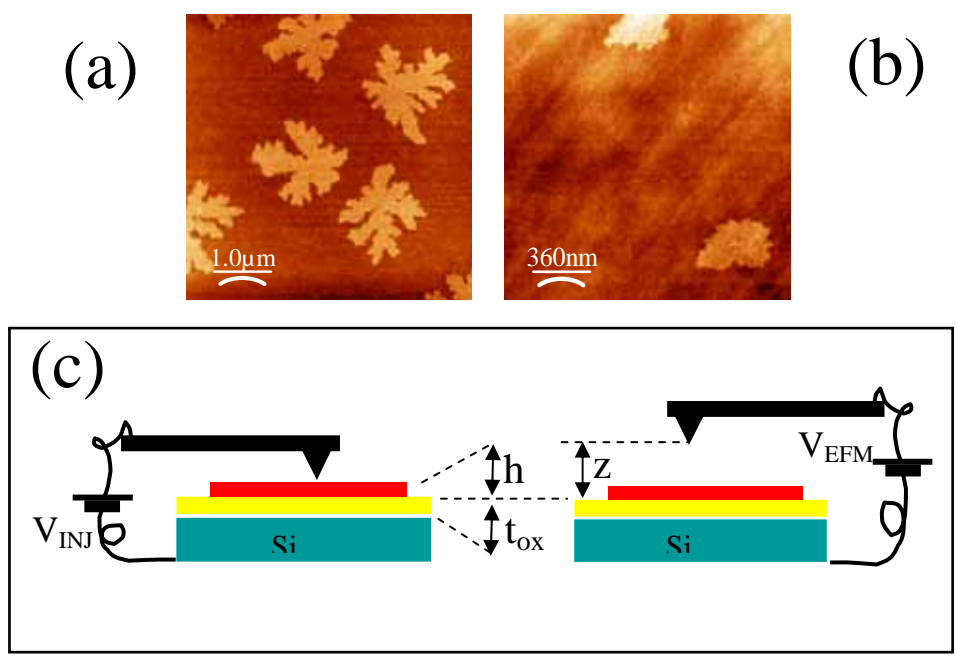

**FIGURE 1**. (a) TM-AFM image (5 µm x 5 µm) of pentacene monolayer islands grown on substrate **I**; (b) TM-AFM image (1.8 µm x 1.8 µm) of pentacene monolayer islands grown on substrate **II**; (c) Schematic representation of charge injection and detection by EFM. The silicon substrate is covered by an oxide (thickness $t_{ox}$). The pentacene island has a height h. In the EFM mode, the tip is set at a distance z from the substrate.

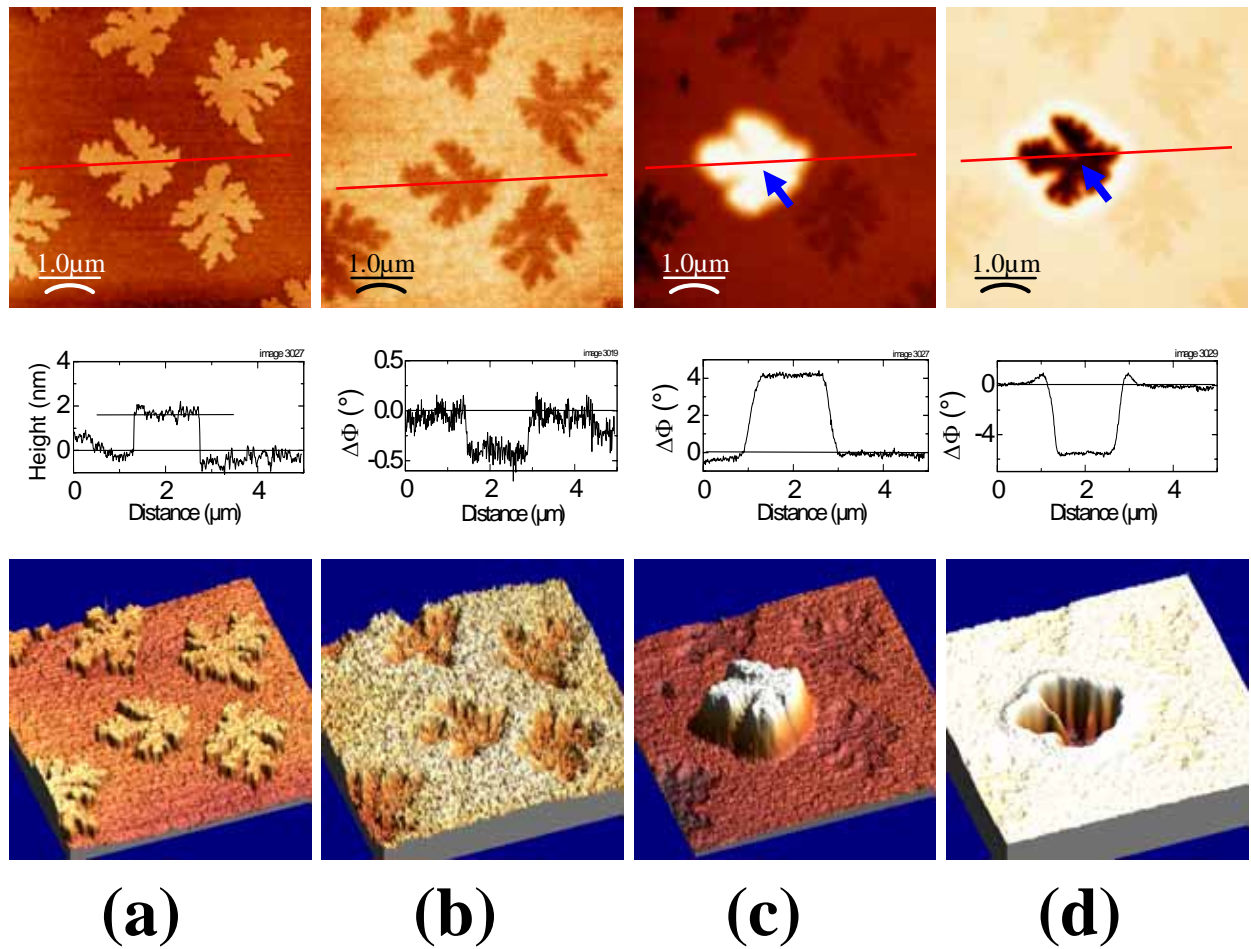

**FIGURE 2.** From top to bottom: 2D images, section profiles along the red line and 3D images of pentacene islands on substrate **I**. (a) TM-AFM of the islands. (b) EFM images (phase shift, $V_{EFM} = 4V$, $z = 80$ nm) of the same islands before charge injection. (c) EFM images (phase shift, $V_{EFM} = 4V$, $z = 80$ nm) of the same islands after a local injection ($V_{INJ} = +2V$ for 4 min) on the central island (injection point marked by the arrow). (d) EFM images (phase shift, $V_{EFM} = 4V$, $z = 80$ nm) after a subsequent local injection at $V_{INJ} = -2V$ for 4 min). All images are 5μm x 5 μm.

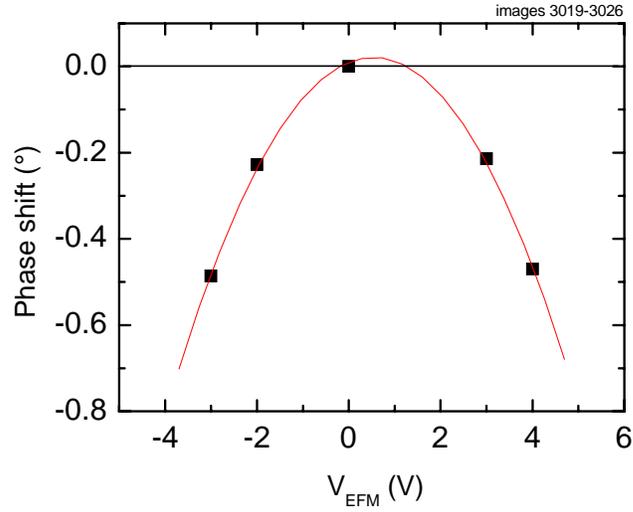

**FIGURE 3**. Phase shift vs. EFM bias ($\Delta\Phi$-$V_{EFM}$) measured before injection on the central island shown in Fig. 2-b. Tip-substrate distance is z = 80 nm. The line is the best fit of a polynomial law $\Delta\Phi = a(V_{EFM} - V_S) + b(V_{EFM} - V_S)^2$, with a = 0.043 °/V, b = -0.04 °/V² and a surface potential $V_S$ = 0.02V. The ratio a/b is used to calculate the residual charge in this pentacene island. The a/b analytical formula is similar to Eq. 1 without the term $V_{EFM}$-$V_S$.

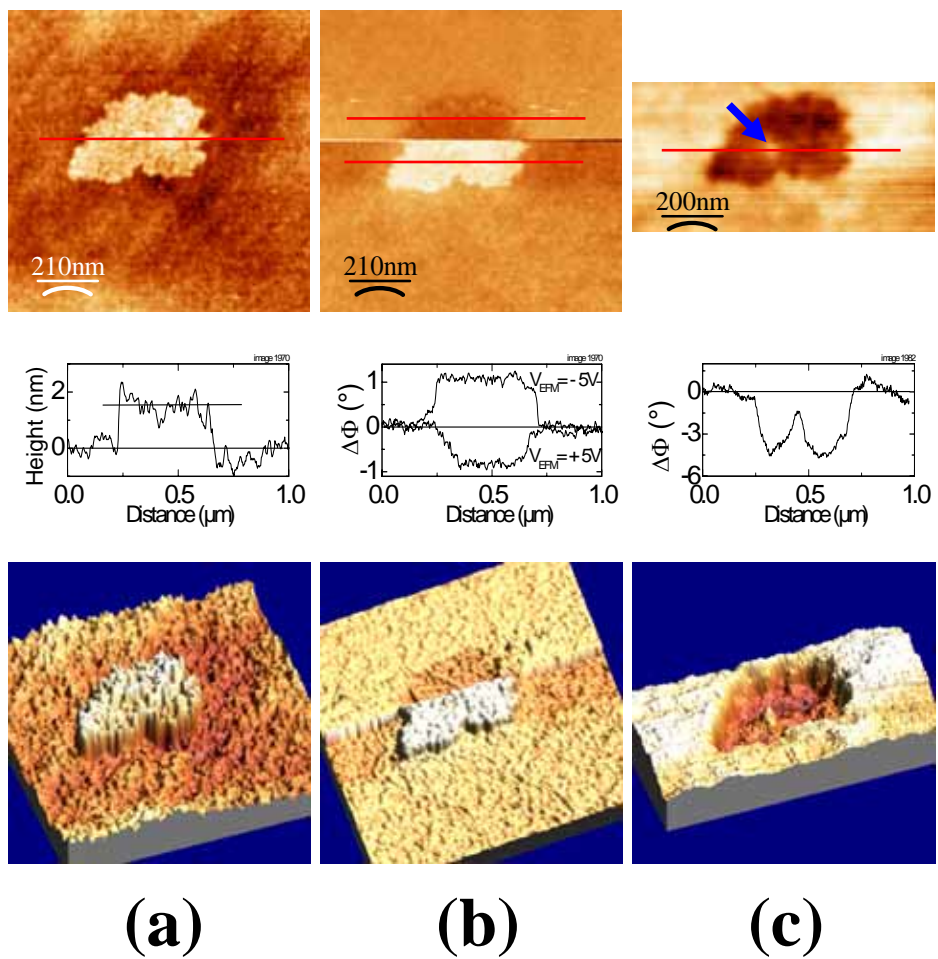

**FIGURE 4.** From top to bottom: 2D images, section profiles along the red line and 3D images of pentacene islands on substrate **II**. (a) TM-AFM images (1 µm x 1 µm) of one island. (b) EFM images (phase shift) of the same island before charge injection (1 µm x 1 µm). The upper half of the image was taken at $V_{EFM}$ = 5V (z = 50 nm), the lower part at $V_{EFM}$ = -5V (z = 50 nm). (c) EFM images (phase shift, $V_{EFM}$ = 5V, z = 20 nm) of the same island after a local injection ($V_{INJ}$ = +5V for 30 s) at the point marked by the arrow.

REFERENCES.

(13) Since the roughness has to be measured on a small area on the island, the roughness values are less accurate than on the bare substrate where larger surface areas can be used to determine the roughness data.

(14) Melin, T.; Diesinger, H.; Deresmes, D.; Stievenard, D. Phys. Rev. B 2004, 69, 035321.

(15) A correction is required because we used a semiconductor substrate instead of a metal. However, this correction can be neglected when using the ratio method, because almost the same correction factor applies in the capacitive and charge gradient forces. See Ref. 14.

(16) According to the calculation of the factor g given in Ref. 14, the increase in the factor g reflects the decrease in the island charge screening when a thin insulating layer is inserted between the pentacene island and the silicon.

(17) Fritz, S. E.; Martin, S. M.; Frisbie, C. D.; Ward, M. D.; Toney, M. F. J. Am. Chem. Soc. 2004, 126, 4084-4085.

(18) Watkins, N. J.; Gao, Y. J. Appl. Phys. 2003, 94, 5782-5786.

(19) Hill, I. G.; Rajagopal, A.; Kahn, A.; Hu, Y. Appl. Phys. Lett. 1998, 73, 662-664.

(20) To our knowledge, the interfacial dipole at the pentacene-Pt/Ir interface has not been measured by photoemission spectroscopy as in other metal/organic systems. However, such a value of about 1 eV is not unreasonable and it has been observed for various interfaces: Au/Alq3 , Au/NPD, Au/TPD and Au/DPNTCI for examples. Alq3=tri(8-hydroxyquinolino)aluminium, NPD=N,N'-diphenyl-N,N'-bis(1-naphthyl)-1,1'-biphenyl-4,4'-diamine, TPD=N,N'-diphenyl-N,N'-(3-methylphenyl)-1,1'-biphenyl-4,4'-diamine, DPNTCI=N,N'-diphenyl-1,4,5,8-naphthyltetracarboxilicimide. See a review in Ishii, H.; Sugiyama, K.; Ito, E.; Seki, K.; Adv. Mat. 1999, 11, 605-625 and references therein.

(21) This may be due to a lower chemical quality of the pentacene batch used for sample II. Chemicals were used as received.

TOC Graphic.

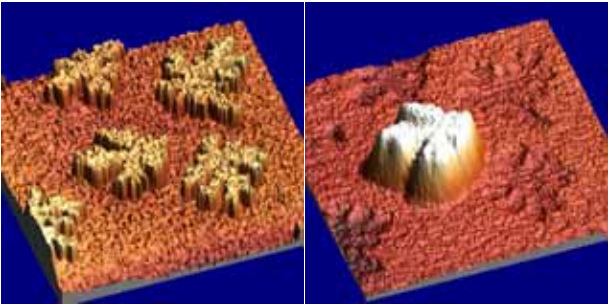